\pdfoutput=1

\documentclass[11pt,twoside,a4paper,cmspaper,final,collab]{cms-tdr}
\def\svnVersion{416586}\def\cmsCernNoTag{CERN-EP/2016-281}\def\cmsCernDate{\today}\def\cmsMessage{Published in Physics Letters B as \href{http://dx.doi.org/10.1016/j.physletb.2017.04.043}{\doi{10.1016/j.physletb.2017.04.043.}}}

\begin{document}\cmsNoteHeader{EXO-15-006}

\hyphenation{had-ron-i-za-tion}
\hyphenation{cal-or-i-me-ter}
\hyphenation{de-vices}
\RCS$HeadURL: svn+ssh://svn.cern.ch/reps/tdr2/papers/EXO-15-006/trunk/EXO-15-006.tex $
\RCS$Id: EXO-15-006.tex 416534 2017-07-17 13:59:53Z delacruz $
\newlength\cmsFigWidth
\ifthenelse{\boolean{cms@external}}{\setlength\cmsFigWidth{\columnwidth}}{\setlength\cmsFigWidth{0.65\textwidth}}
\ifthenelse{\boolean{cms@external}}{\providecommand{\cmsLeft}{upper\xspace}}{\providecommand{\cmsLeft}{left\xspace}}
\ifthenelse{\boolean{cms@external}}{\providecommand{\cmsRight}{lower\xspace}}{\providecommand{\cmsRight}{right\xspace}}
\ifthenelse{\boolean{cms@external}}{\providecommand{\cmsULeft}{upper\xspace}}{\providecommand{\cmsULeft}{upper left\xspace}}
\ifthenelse{\boolean{cms@external}}{\providecommand{\cmsURight}{middle\xspace}}{\providecommand{\cmsURight}{upper right\xspace}}

\providecommand{\Wprime}{\PWpr}
\providecommand{\MT}{\ensuremath{M_\mathrm{T}}\xspace}
\providecommand{\MTlower}{\ensuremath{M_\mathrm{T}^\mathrm{min}}\xspace}
\providecommand{\MPT}{\ensuremath{\vec{p}_{\text{T}}^{\text{miss}}}\xspace}
\providecommand{\pb}{\ensuremath{\unit{pb}}\xspace}
\providecommand{\fb}{\ensuremath{\unit{fb}}\xspace}

\cmsNoteHeader{EXO-15-006}
\title{Search for heavy gauge \Wprime bosons in events with an energetic lepton and large missing transverse momentum at $\sqrt{s} = 13\TeV$}

\date{\today}

\abstract{
A search is presented for \Wprime bosons in events with an electron or muon and large missing transverse momentum, using proton-proton collision data at $\sqrt{s} = 13\TeV$
collected with the CMS detector in
2015 and corresponding to an integrated luminosity of 2.3\fbinv.
No evidence of an excess of events relative to the standard model expectations is observed.
For a \Wprime boson described by the sequential standard model, upper limits at 95\% confidence level are
set on the product of the production cross section and branching fraction and lower limits are established on the
new boson mass. Masses below 4.1\TeV are excluded
combining electron and muon decay channels,
significantly improving upon the results obtained with the 8\TeV data. Exclusion limits at 95\% confidence level on the product of the \Wprime production cross section and branching fraction are also derived in combination with the 8\TeV data. Finally,
exclusion limits are set for the production of generic \Wprime bosons decaying into this final state using a model-independent approach.
}

\hypersetup{%
pdfauthor={CMS Collaboration},%
pdftitle={Search for heavy gauge W' boson in events with an energetic lepton and large missing transverse momentum at sqrt(s) = 13 TeV},%
pdfsubject={CMS},%
pdfkeywords={CMS, physics, Exo, BSM searches, Wprime}}

\maketitle

\section{Introduction}
\label{sec:intro}

Many standard model (SM) extensions require additional heavy gauge bosons. In particular, the sequential standard model (SSM)~\cite{reference-model} predicts the existence of a new massive boson, \Wprime,
exhibiting the same couplings as the SM W boson, i.e., with final states consisting either of a charged lepton and neutrino or a quark pair.
If sufficiently massive, the decay channel $\Wprime\to \PQt\PAQb$ is also allowed.

\begin{figure}[hbtp]
\centering
\includegraphics[width=.55\textwidth]{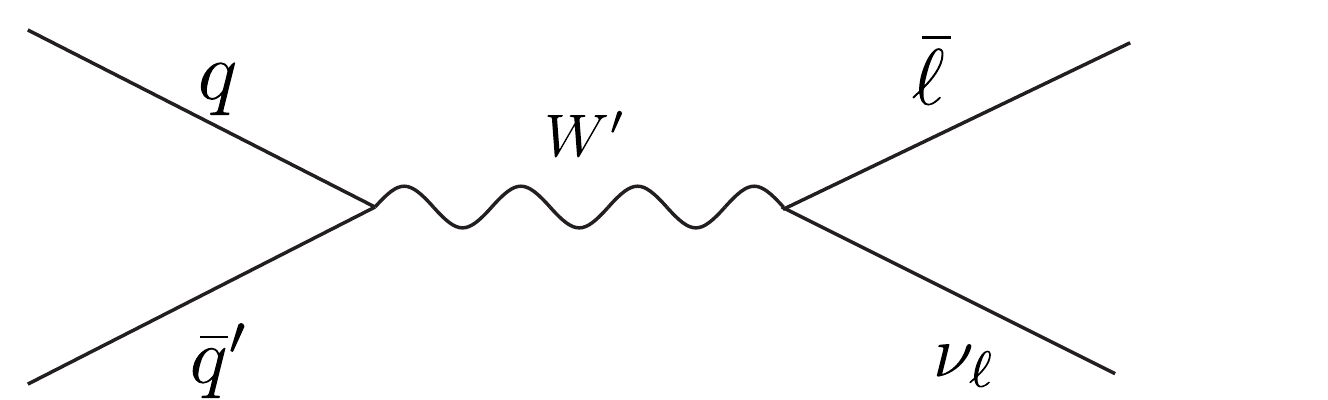}
\caption{Production and decay of an SSM \Wprime boson. The final state shown denotes both the
($\overline{\ell}\nu_{\ell}$) state and its charge conjugate.}
\label{fig:feynman}
\end{figure}

This Letter describes a search for deviations from the SM predictions in events with a
charged lepton (electron or muon) and missing transverse momentum in the final state, proceeding as shown in Fig.~\ref{fig:feynman}.
It is assumed that there is no interference between the production of the new particle and the production of the SM W boson.  This would be the case, for example, if the \Wprime interacts via V+A coupling.
Its decays to SM bosons (W, Z, H), which are model dependent, are neglected.
Dedicated searches for \Wprime decays into bosons can be found in Refs.~\cite{WprimeWZ1, WprimeWZ2, AtlasDiboson}.

Similar searches have been carried out by experiments at the FNAL Tevatron~\cite{D0_Wprime, CDF_Wprime}.
The most stringent limits on the mass of an SSM \Wprime boson to date come from the CERN LHC experiments.
Using an integrated luminosity of 19.7\fbinv of proton-proton (pp) collisions at a center-of-mass energy of 8\TeV, CMS set a lower limit at 95\% confidence level (CL) on the \Wprime boson mass of 3.22\TeV in the
electron channel and 2.99\TeV in the muon channel~\cite{EXO-12-060}.
Combining both channels resulted in an exclusion of \Wprime bosons with a
mass less than 3.28\TeV.
Similarly, for the combined channels at $\sqrt{s}=8\TeV$, ATLAS excluded \Wprime bosons with masses less than 3.24\TeV~\cite{Wprime_ATLAS}.

Because of the increase in the center-of-mass energy from 8 to 13\TeV, the parton luminosities associated with $\qqbar'$ interactions producing the \Wprime bosons increase by more than an
order of magnitude in the high-mass region. Limits derived by ATLAS~\cite{Wprime_ATLAS13TeV} using 3.2\fbinv of pp collisions at $\sqrt{s}=13\TeV$ exclude SSM \Wprime bosons with masses less than
4.07\TeV, for the combination of the electron and muon decay channels.

The results presented in this Letter are based on the analysis of 2.3\fbinv of pp collision data collected with the CMS detector during 2015, at $\sqrt{s}=13\TeV$.
Limits are given both for the SSM interpretation, and
for a generic \Wprime,
enabling constraints to be placed on a variety of other models.

\section{The CMS detector}\label{sec:detector}

A detailed description of the CMS detector and the coordinate system used can be found in Ref.~\cite{refCMS}.
The central feature of the CMS apparatus is a superconducting
solenoid of 6\unit{m} internal diameter, providing a magnetic field of 3.8\unit{T}.
Within the solenoid volume
are located the silicon pixel and strip tracker, measuring charged-particle trajectories in the pseudorapidity region $\abs{\eta} < 2.5$, and the barrel and two endcap sections of the calorimeters ($\abs{\eta}<3$): a lead tungstate crystal electromagnetic calorimeter (ECAL), and a brass and scintillator hadron calorimeter (HCAL).
Forward calorimeters extend the $\eta$ coverage provided by the barrel and endcap detectors ($3<\abs{\eta}<5$). Muons are measured in gas-ionization detectors embedded in the steel flux-return yoke
outside the solenoid, in the range $\abs{\eta} < 2.4$, with detection planes made using three technologies: drift tubes, cathode strip chambers, and resistive plate chambers. Additional detectors and upgraded electronics, installed before the beginning of the 13\TeV data collection period in 2015, have yielded improved reconstruction performance for muons relative to the 8\TeV data collection period in 2012.

The CMS experiment has a two-level trigger system. The level-1 (L1) trigger~\cite{TriDASI}, composed of
custom hardware processors, selects events of interest using information from the calorimeters
and muon detectors and reduces the readout rate from the 40\unit{MHz} bunch-crossing frequency
to a maximum of 100\unit{kHz}. The software based high-level trigger (HLT)~\cite{TriDASII} uses the full event
information, including that from the inner tracker, to reduce the event rate to the 1\unit{kHz} that is
recorded.

\section{Analysis strategy and simulated samples}
\label{sec:models}

The analysis selects events with
a high-energy charged lepton and large missing transverse
momentum (\ptvecmiss),
which may indicate the presence of a non-interacting particle (neutrino).
The quantity \MPT is defined as ${-}\sum \ptvec$ of all reconstructed particles with \MET being the magnitude of \MPT.

The main discriminating variable used in the search is the transverse mass
defined as
$\MT = \sqrt{\smash[b]{2 \pt^\ell\ETm (1-\cos[\Delta\phi(\ptvec^\ell,\MPT)])}}$, where
$\ptvec^\ell$ is the lepton transverse momentum, $\pt^\ell$ is its magnitude,
and $\Delta\phi(\ptvec^\ell,\MPT)$ is the difference in azimuthal angle between the lepton transverse momentum and missing transverse momentum vectors.
A signal from very massive \Wprime bosons would appear at high \MT values.

The dominant and irreducible background is W $\to \ell \nu$ with $\ell= \Pe,\mu, \tau$. The $\PW\to \tau \nu$ process mostly contributes to the
region of lower \MT values relative to decays into the other lepton channels, because of the momenta carried away by the two neutrinos from the tau decay.
Possible interference between the production of \Wprime and SM W bosons is not considered. The existance of interference effects would change the total cross section and the shape of the \MT spectrum~\cite{EXO-12-060}.
Other background processes are Drell--Yan (where one of the leptons is not reconstructed),
\ttbar (semileptonic and dileptonic decay channels), single top quark, and dibosons (mainly WW production). The contributions from these processes are estimated from simulation.

To estimate the dominant SM W boson background, different
$\PW\to \ell \nu$ samples are used: an inclusive one generated at next-to-leading order (NLO) with \MADGRAPH5\_a\MCATNLO~\cite{Madgraph} describing the events with a W boson mass up to 200\GeV, and several exclusive samples, covering the
boson high-mass region (from 200\GeV onwards), generated at leading order (LO) with \PYTHIA~8.2
~\cite{Sjostrand:2014zea},
tune CUETP8M1~\cite{CUETP8M1,GEN-14-001-PAS}, and NNPDF3.0 parton distribution functions (PDF)~\cite{Ball:2014uwa}.
A mass-dependent $K$ factor, to account for higher order effects, is calculated using {\FEWZ}~$3.1$~\cite{Li:2012wna}
at next-to-next-to-leading order (NNLO) QCD precision and {\sc mcsanc}~1.01~\cite{Bondarenko:2013nu}
at NLO electroweak precision.
The application of the $K$ factor
improves the description of the tail of the \MT distribution, the key element in this search.

High mass Drell--Yan and \ttbar samples are generated with \POWHEG(v2)
~\cite{Powheg, Powheg_v2,Alioli:2009je,Re:2010bp,Campbell:2014kua}, an event generator at NLO, with parton showering and hadronization described by \PYTHIA8.2, using the
\newline CUETP8M1 tune and NNPDF3.0 PDF set.
The \ttbar category includes both semileptonic and dileptonic decay modes samples.
Single top quark production is generated inclusively with \POWHEG(v2) in the tW-channel and with \MADGRAPH5\_aMC@NLO matched to \PYTHIA8.2 using the FXFX algorithm~\cite{FXFXmerging}, in the $s$- and
$t$-channels.
Diboson (WW, WZ, and ZZ) production is generated with \PYTHIA8.2, \mbox{tune CUETP8M1}, and the NNPDF2.3LO PDF set~\cite{NNPDF23}.

Background from jets misidentified as electrons (referred to as QCD multijet background in what follows) is largely rejected
by the analysis selection criteria described in the next section, and the residual
contribution is estimated from data by using
a control region defined by the electron isolation and the ratio $\pt^\ell/\MET$.
This method of estimating the QCD multijet contribution was already used in our previous analysis~\cite{EXO-12-060}
and is based on four regions (isolated and non-isolated signal and background events) to estimate the normalization and provide the template data.
The probability to misidentify jets as muons is negligible.

For the signal events, the generation of SSM \Wprime $\to \ell\nu$ samples for the electron and muon decay channels is performed with \PYTHIA8.2 at LO,
tune CUETP8M1, and the NNPDF3.0 PDF set.
A \Wprime mass-dependent $K$ factor is applied based on NNLO QCD cross sections as calculated with {\FEWZ}~$3.1$.
The $K$ factor varies from 1.3 to 1.1 for the range of \Wprime masses studied in this analysis, namely from 0.4 to 5.8\TeV.
The NNLO corrections decrease with \Wprime boson masses up to around 4.5\TeV.
For higher \Wprime masses, the phase space for production in pp collisions at 13\TeV decreases, leading to a growing fraction of new bosons produced off mass-shell, towards lower masses.
In those cases, the $K$ factor increases and becomes similar to the low-mass values.
The product of the NNLO signal production cross section and branching fraction, $\sigma_{\Wprime}
 \mathcal{B}(\Wprime \to \ell \cPgn)$, with $\ell = \Pe$ or $\mu$, strongly depends on the \Wprime mass, varying from 111\pb for M(\Wprime) = 0.4\TeV to 0.151\fb for M(\Wprime) = 5.8\TeV. For the benchmark masses of
M(\Wprime) = 2.4 and 3.6\TeV, used later for illustration, the values are 59.8 and 4.4\fb, respectively.
The width of the SSM \Wprime is a function of its mass.

All generated signal and background events are processed through a full simulation of the CMS detector based on \GEANTfour~\cite{Agostinelli:2002hh}, and including an emulation of the trigger. The simulated events are reconstructed with the same code used to reconstruct the data.

The simulation of particle production from additional collisions in the same or nearby bunch crossing (pileup) is included in all event samples by superimposing minimum bias interactions onto the simulated
events, with a frequency distribution matching that observed in data.
The average number of interactions per bunch crossing in the selected data is 10.

\section{Object identification and event selection}
\label{sec:selection}

Events with at least one high-\pt lepton are selected using inclusive lepton triggers.
Single-electron triggers with transverse energy thresholds of
105 or 115\GeV and loose electron identification criteria are used. The single-muon triggers require $\pt > 45\GeV$ for a muon pseudorapidity, $\abs{\eta} < 2.1$, or $\pt >
50\GeV$ for $\abs{\eta} < 2.4$ (the limit of coverage of the muon detectors).
The relatively high electron trigger threshold is required in order to suppress non-prompt electrons and misidentified jets. The offline reconstructed \pt must be greater than 130 (53)\GeV in the electron (muon) channel, where the trigger efficiency reaches a plateau of 0.99 (0.96) relative to the full analysis requirements described in the following.

Leptons and \ptvecmiss
are reconstructed using a particle-flow technique~\cite{PF1, PF2}, an algorithm that combines measurements from all components of the CMS detector in order to reconstruct and identify
individual particles in the event. Requirements for identifying good quality and energetic leptons are applied, optimized for high-\pt values
where the analysis has the largest sensitivity to the expected signals.
Events containing calorimeter noise or large \MET due to instrumental effects, such as beam halo or jets near nonfunctioning channels in the calorimeters~\cite{CMS-PAS-JME-16-004}, are not used.
The primary vertex in the event is defined as the vertex with the highest $\sum \pt^2$, where the sum is over the tracks associated to it.

Electrons are reconstructed from electromagnetic energy deposits (clusters) in the ECAL acceptance region (barrel, $\abs{\eta} < 1.444$, endcaps, $1.566 < \abs{\eta} < 2.5$) matched to a track
in the silicon tracker~\cite{EGM-13-001}. The transverse energy of a localized cluster is defined as $\ET = E \sin\theta$, with $\theta$ the polar angle relative to the beam axis, and where the cluster energy $E$ includes any deposits consistent with bremsstrahlung emission.
The electron identification, optimized for high-\pt values~\cite{EXO-12-061,EXO-15-005}, includes requirements on the isolation and on the variables describing the electromagnetic shower shape.
The electron isolation is computed using the sum of three terms, based on tracker, ECAL, and HCAL information, after correction for the contributions expected from detector noise and pileup.
 The electron isolation in the tracker is ensured by requiring the scalar \pt sum of all tracks, within a cone of $\Delta R = \sqrt {\smash[b]{(\Delta\eta)^2+(\Delta\phi)^2}} = 0.3$ centered around the track of the electron candidate and originating from the primary vertex, to be less than
5\GeV.
The ECAL isolation is defined as the \ET sum of the energy deposits within a cone of $\Delta R = 0.3$ around the electron candidate to be less
than 3\% of the electron \ET. The HCAL isolation considers the sum of energy deposits in the hadronic calorimeter within a cone of $\Delta R = 0.15$ around the electron direction which must be less than 5\% of the electron energy deposit in the ECAL. In each case the sums exclude the electron candidate itself.
In order to differentiate between electrons and photon conversions, the electron track is required to have no more than one hit missing in the pixel layers, and the transverse distance to the primary vertex must be less than 0.02 (0.05)\cm in the barrel (endcap).
The electron momenta for electrons with $\pt \approx 45$\GeV from $\Z \to \Pe \Pe$ decays are estimated by combining energy measurements in the ECAL with momentum measurements in the tracker. For high-energy electrons the momentum scale and resolution are dominated entirely by the energy measurement in the ECAL. The discriminating variable in this search, \MT, is more sensitive to variations of energy scale than to uncertainty in energy resolution.  The energy scale has therefore been checked using high-mass offshell dielectron events coming from Z-boson decays.

 Muons are reconstructed by combining the information from the tracker and the muon systems~\cite{MUO-10-004, MUO-11-001}. Each muon is required to have at least one hit in the pixel detector, hits in at least four layers of the strip tracker, and segments in two or more muon detector chambers. Since consecutive layers in the muon system are separated by thick
layers of steel, the latter requirement significantly reduces the amount of hadronic \textit{punch-through}. To reduce
background from cosmic ray muons, each muon is required to have a transverse impact parameter
less than 0.02\cm and a longitudinal distance parameter
less than 0.5\cm. Both parameters are defined relative to the primary vertex. In order to suppress muons with mismeasured $\pt$, an additional requirement $\sigma_{\pt}/\pt < 0.3$ is applied, where $\sigma_{\pt}$ is the
uncertainty in the \pt from the track reconstruction.
Muon isolation requires that the scalar $\pt$ sum of all tracks originating from the interaction vertex within a cone of $\Delta R = 0.3$ around its direction, excluding the muon track, be less than 10\% of the muon $\pt$.
The muon \pt reconstruction is optimized for the high-\pt region and its performance  has been studied using both high-energy cosmic ray muons and dimuons from high-\pt \PZ boson decays~\cite{EXO-15-005}. The relative \pt resolution for muons with $\pt < 200\GeV$ is 1.3--2.0\% in the barrel and better than 6\% in the endcaps. For high-\pt muons (\pt up to 1\TeV) the relative resolution in the barrel is better than 10\%.

To reduce the Drell--Yan background in each decay channel, events with additional electrons (muons) of $\pt > 35$ (25)\GeV and in $\abs{\eta} < 2.5 (2.4)$ are rejected.

Once events containing a high-\pt lepton are selected, the two-body decay kinematics of the \Wprime$\to \ell \nu$ process is exploited to further reduce the background, by applying two additional requirements, $\abs{\Delta\phi(\ptvec^\ell, \MPT)} > 2.5$ and $0.4 < \pt^\ell/\MET < 1.5$.

The signal efficiency for the selection procedure, with no requirement on the reconstructed \MT in the event, is determined from simulated samples and is maximal ($\approx$0.80 for both decay channels) for a \Wprime boson of mass 1.5\TeV and decreases gradually for larger and smaller masses down to $\approx0.65$.

\section{Systematic uncertainties}
\label{sec:uncertainties}

The sources of systematic uncertainties of experimental nature can be divided into those that are channel-specific and those that are common to the electron and muon channels.
For each source of uncertainty, upper and lower values are propagated to the kinematic quantities of the different objects (e, $\mu$, and \MET) in each event, the selection re-applied, and new \MT values obtained, which are considered in the statistical analysis of the data, as presented in the next section.

Mismeasurements of the electron energy scale and resolution are typically small and do
not change the \MT shape significantly. The systematic uncertainty in the electron energy scale was
taken to be 2\%~\cite{EXO-15-005}.
For the electron energy resolution, an additional Gaussian smearing of 2\% is applied to the one from
MC simulation, to match the measurements performed on data using dielectron events from \PZ boson decays.
The uncertainty in the electron identification efficiency
when extrapolated to high \ET
is 4\% (6\%) in the barrel (endcaps).
Scale factors are applied to the simulation samples to account for possible differences between data and simulation for trigger efficiency. They are determined with an uncertainty of 0.2\% (0.5\%) for barrel (endcaps), and are consistent with unity for the electrons.

In the muon decay channel,
the \pt scale is sensitive to an imperfect modeling of the alignment in the tracker or muon systems.
Studies are performed on the curvature of muon tracks in different regions of $\eta$ and $\phi$ using high-\pt cosmic ray data and dimuon events from collisions, together with the corresponding simulation samples. They indicate the absence of a significant curvature bias.
The uncertainties associated with these results are taken as contributions to the overall systematic uncertainties.
For the central region ($\abs{\eta} <$ 1.2) the bias uncertainty is 0.03/TeV and in the forward region (1.2 $< \abs{\eta} <$ 2.4) the bias uncertainty is 0.04/TeV. These uncertainties are propagated to the muon \pt assignment and consequently, to the \MT distribution.
The \pt resolution at high-\pt values in data is well reproduced by the simulation and no further correction is applied.
Muon trigger and identification efficiencies measured in data are consistent with those from simulated samples within the precision of the efficiency measurement allowed by the amount of data collected at high \pt. Uncertainties on the extrapolation to high \pt values are assigned, which increase from 3\% for $\pt < 500\GeV$ to 8.5\% for $\pt > 1\TeV$.

The sources of uncertainty in the lepton \pt translate directly into the \MET calculation, which in the sample of events selected is mainly determined by the high \pt of the lepton. As
events are allowed to include an arbitrary number of jets, which may originate from
initial state radiation, systematic uncertainties in the
jet energy scale and resolution are propagated to the $\MET$ variable.

Common to both the electron and muon channels are the uncertainties on the total integrated luminosity (2.7\%)~\cite{CMS-Lumi} and in the reweighting procedure applied to simulated samples to match the pileup in data (5\%).
The application of $K$ factors accounting for
higher-order corrections, both for the signals and
the background, is also affected by systematic uncertainties.
The uncertainty in the signal $K$ factor arises from the choice of PDF and $\alpha_S$. The combined uncertainty is evaluated using the PDF4LHC prescription~\cite{PDF4LHCatRUNII}, where in the computation of each PDF set the strong coupling constant is varied. Uncertainties from different PDF sets and $\alpha_S$ variation are added in quadrature.
For the background $K$ factor, a uniform uncertainty of 5\%, stemming from the NNLO corrections,
is applied in addition to a mass-dependent uncertainty. The latter is determined
by comparing the results from the two possible procedures for combining the
QCD and electroweak corrections: additive or factorized methods~\cite{EXO-12-060}.
The theoretical uncertainty related to the choice of the PDF
set in the background modeling is estimated using the PDF4LHC prescription and
dominates the total uncertainty at high \MT in both decay channels.

\section{Results}
\label{sec:results}

\begin{figure}[hbtp]
\centering
 \includegraphics[width=0.49\textwidth]{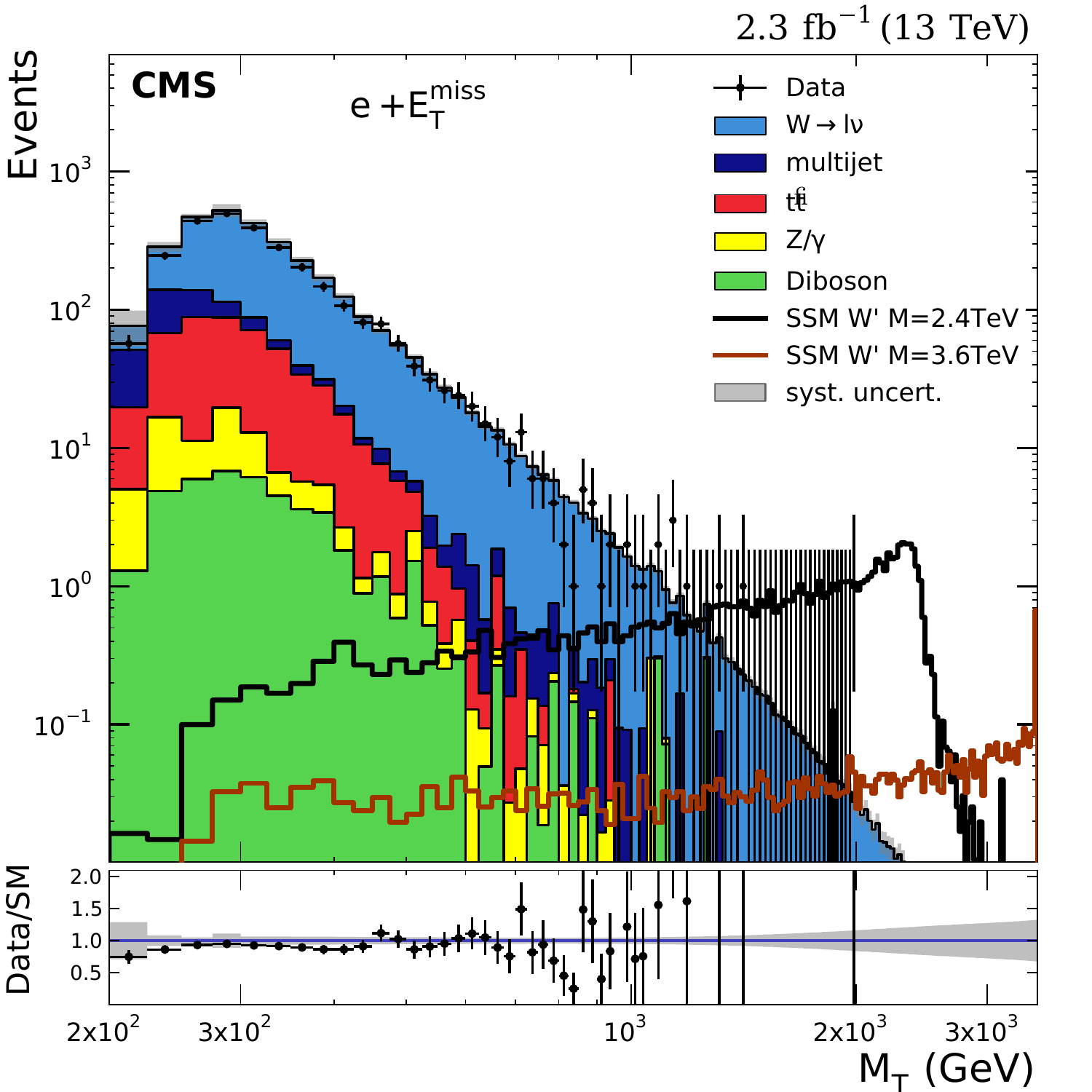}
 \includegraphics[width=0.49\textwidth]{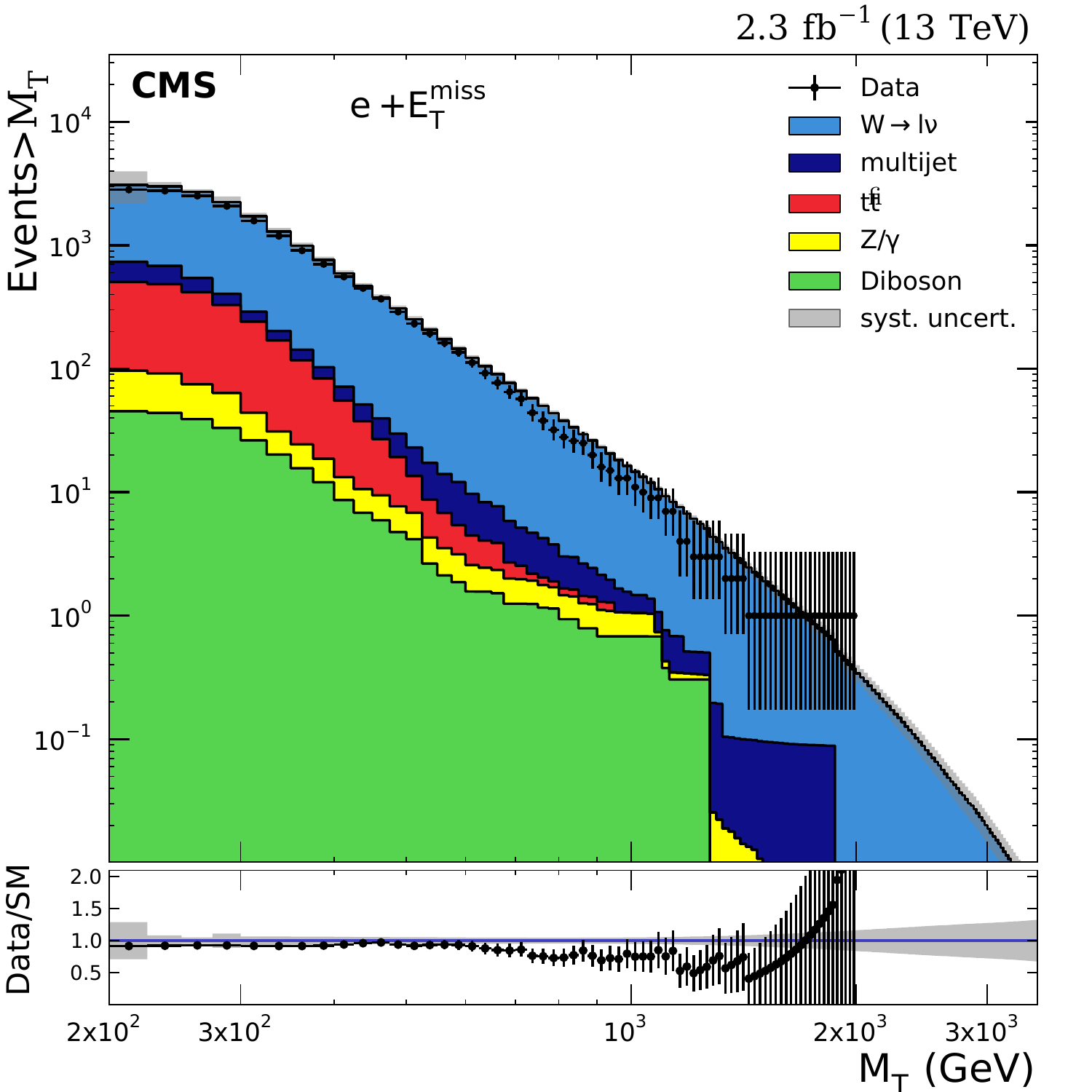}
 \caption{Distributions for data and expected SM backgrounds in the electron channel:
 transverse mass \MT (\cmsLeft) and the associated integral distribution (total number of events above a given value of \MT) (\cmsRight).
The expected signals from the decays of \Wprime
bosons with masses $M(\Wprime) = 2.4$ and 3.6\TeV are also shown in the \cmsLeft figure.
The lower panels show the ratio of data to SM predictions, where the band centered around unity indicates the systematic uncertainty in the expected background, excluding the 2.7\% uncertainty in the
luminosity.}
  \label{fig:Distributions-fin-Ele}
\end{figure}

\begin{figure}[hbtp]
\centering
 \includegraphics[width=0.49\textwidth]{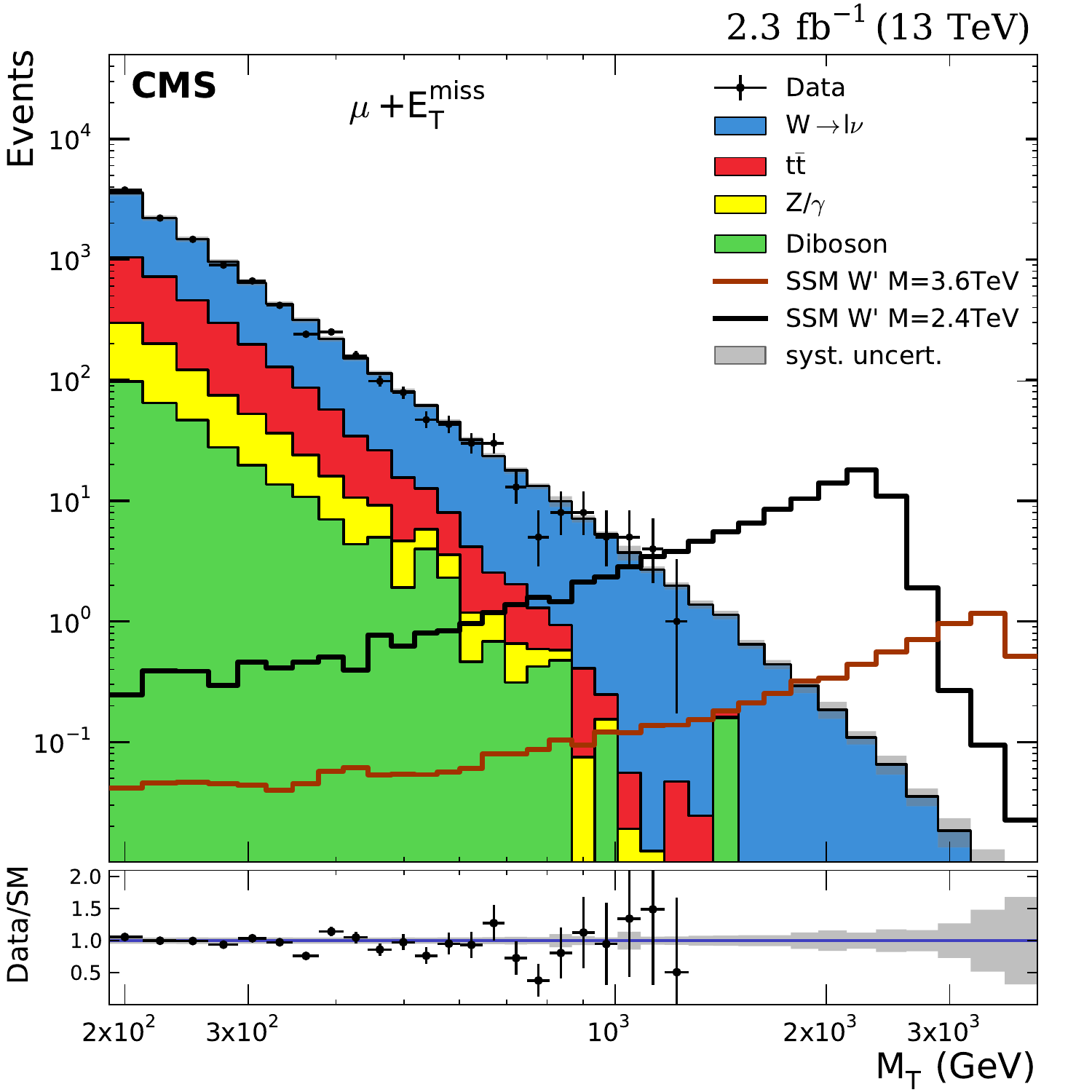}
 \includegraphics[width=0.49\textwidth]{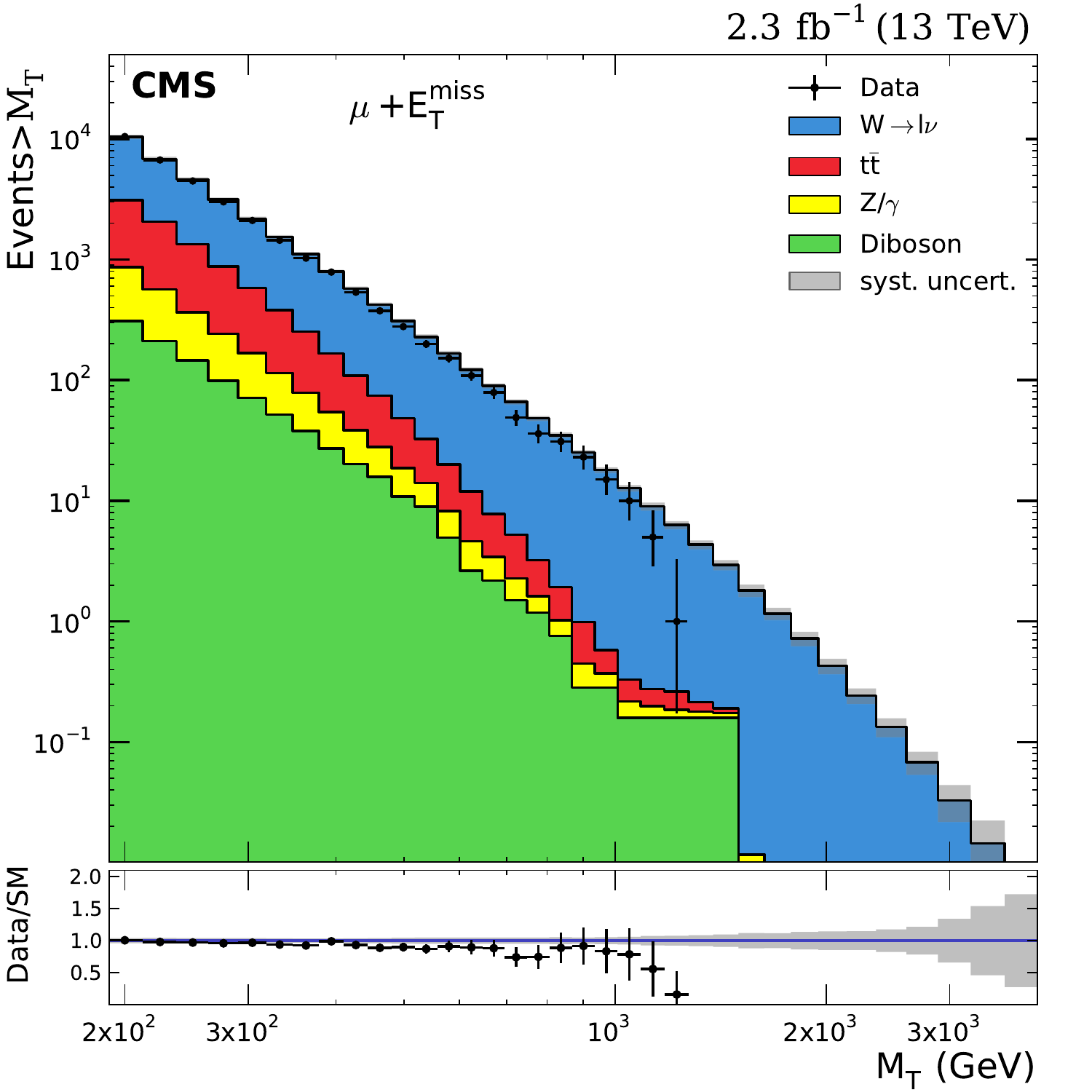}
 \caption{Distributions for data and expected SM backgrounds in the muon channel:
transverse mass \MT (\cmsLeft) and the associated integral distribution (total number of events above a given value of \MT) (\cmsRight).
The expected signals from the decays of \Wprime
bosons with masses $M(\Wprime) = 2.4$ and 3.6\TeV are also shown in the \cmsLeft figure.
The lower panels show the ratio of data to SM predictions, where the band centered around unity indicates the systematic uncertainty in the expected background, excluding the 2.7\% uncertainty in the luminosity.
}
  \label{fig:Distributions-fin-Mu}
\end{figure}

Figure~\ref{fig:Distributions-fin-Ele} shows the distribution of transverse mass \MT (\cmsLeft) and the associated integral distribution (total number of events above a given value of \MT) (\cmsRight)
 for the electron decay channel for $\MT > 200\GeV$. The corresponding distributions are presented for the muon channel in
Fig.~\ref{fig:Distributions-fin-Mu} for $\MT > 120\GeV$,
where the lower trigger \pt threshold enables the extension of the distribution to lower transverse masses.
The increasing bin size at high \MT values in the muon distribution
reflects the degrading muon \pt resolution.
The highest \MT value observed in the electron (muon) channel is 2.0 (1.2)\TeV.
The expected signals from the decay of \Wprime bosons with masses M(\Wprime) = 2.4 and 3.6\TeV are also shown.
The lower panels in the \MT distributions present the data-to-prediction ratios and indicate reasonable agreement between data and SM expectations.

Tables~\ref{tab:ele_event_count} and \ref{tab:muon_event_count} summarize the number of events expected from SM processes, compared to data, when integrating above three representative \MT thresholds (500, 1000, and 1500\GeV)
for the electron and muon decay channels, respectively. Also shown are the number of expected signal events for \Wprime signals with mass M(\Wprime) = 2.4 and 3.6\TeV.

\begin{table*}[hbtp]
\centering
\topcaption{Numbers of events in the electron decay channel integrated above \MT thresholds of 500,
1000, and 1500\GeV, for data, SM background, and signals with M$(\Wprime) =$ 2.4 and 3.6\TeV.
The uncertainties include systematic and statistical uncertainties, but do not include the 2.7\% uncertainty in the integrated luminosity.}
  \label{tab:ele_event_count}
\begin{tabular}{l l l l l} \hline
  &   & $\MT > 500$\GeV & $\MT > 1000$\GeV & $\MT > 1500$\GeV\\\hline
Data &   & 230 & 11 & 1\\\hline
SM Background &   & $246\pm18$ & $14.3\pm1.2$ & $1.9\pm0.2$\\
SSM $\Wprime = 2.4\TeV$ &   & $66.1\pm5.5$ & $58.4\pm5.2$ & $46.3\pm4.4$\\
SSM $\Wprime = 3.6\TeV$ &   & $5.5\pm0.7$ & $4.9\pm0.7$ & $4.3\pm0.6$\\
\hline \end{tabular}
\end{table*}

\begin{table*}[hbtp]
\centering
 \topcaption{Numbers of events in the muon decay channel integrated above \MT thresholds of 500, 1000,
and 1500\GeV, for data, SM background, and signals with M$(\Wprime)$ = 2.4 and 3.6\TeV. The
uncertainties include statistical and systematic uncertainties, but do not include the 2.7\% uncertainty in the integrated luminosity.}
  \label{tab:muon_event_count}
\begin{tabular}{l l l l l}\hline
  &   & $\MT > 500$\GeV & $\MT > 1000$\GeV & $\MT > 1500$\GeV\\  \hline
Data &   & 229 & 11 & 0  \\ \hline
SM Background &   & $255 \pm 20 $ & $12.8 \pm 1.2$ & $1.8 \pm 0.2$\\
SSM \Wprime $M = 2.4\TeV$ &   & $ 95.1 \pm 5.2 $ & $83.2 \pm 5.0 $ & $64.1 \pm 6.0$\\
SSM \Wprime $M = 3.6\TeV$ &   & $ 6.4 \pm 0.5$ & $5.7 \pm 0.5$ & $5.0 \pm 0.5$\\\hline
\end{tabular}
\end{table*}

\subsection{Exclusion limits on SSM \texorpdfstring{\Wprime}{W'} bosons}
\label{sec:results-ssm}

Upper limits on the product $\sigma_{\Wprime}  \mathcal{B}(\Wprime \to \ell \cPgn)$, with $\ell = \Pe$ or $\mu$,
are determined using a Bayesian approach with
a uniform prior probability distribution for the signal cross section in the context of SSM \Wprime boson
production~\cite{CowanPDG}.
A shape analysis (binned likelihood) is performed where the likelihood function is based on probability density functions described by
the \MT distributions for the expected background processes, signals, and data.
Systematic uncertainties discussed in Section~\ref{sec:uncertainties}
in the expected signal and background yields are included through nuisance parameters with log-normal
prior distributions.

\begin{figure}[hbt]
\centering
 \includegraphics[width=0.49\textwidth]{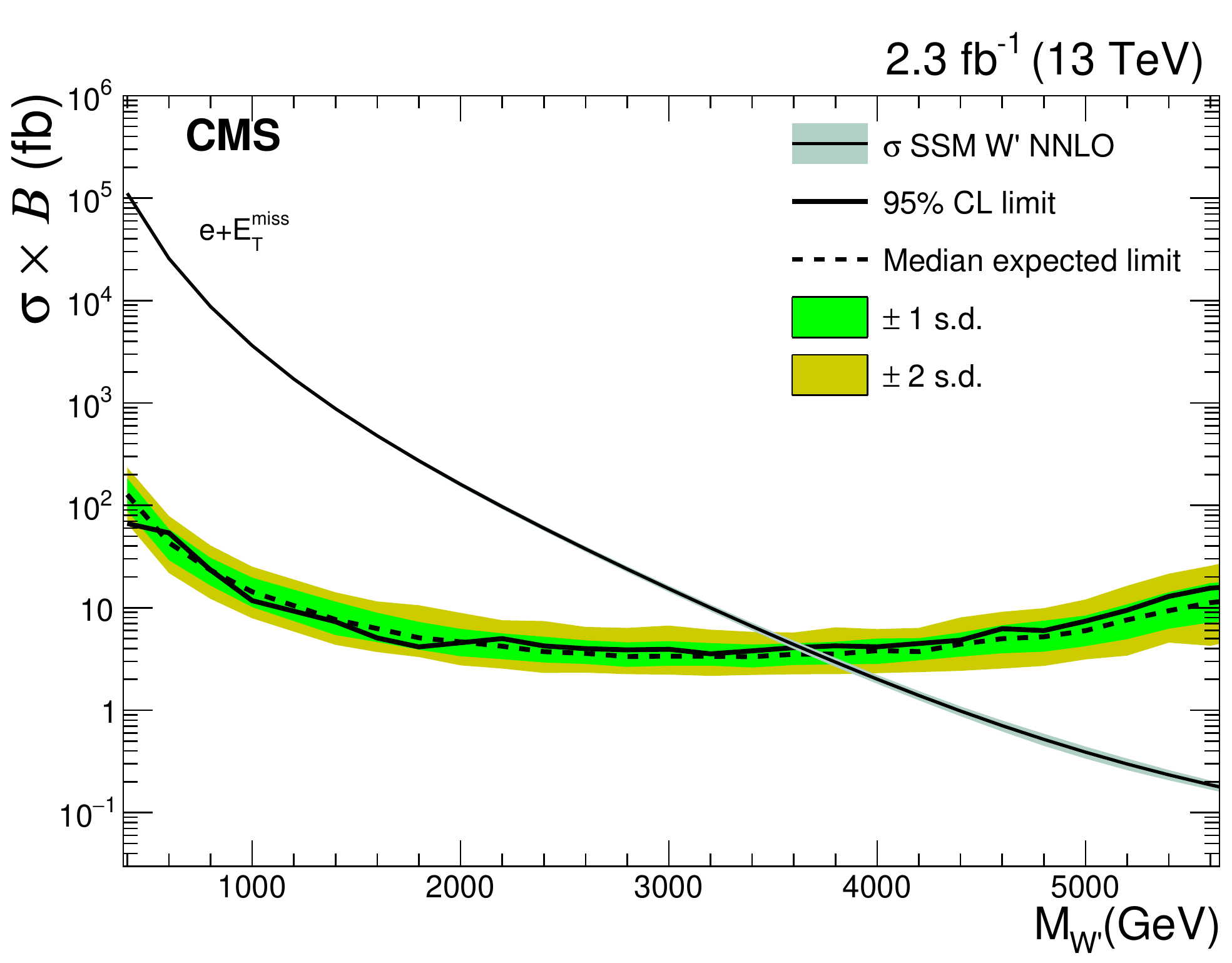}
 \includegraphics[width=0.49\textwidth]{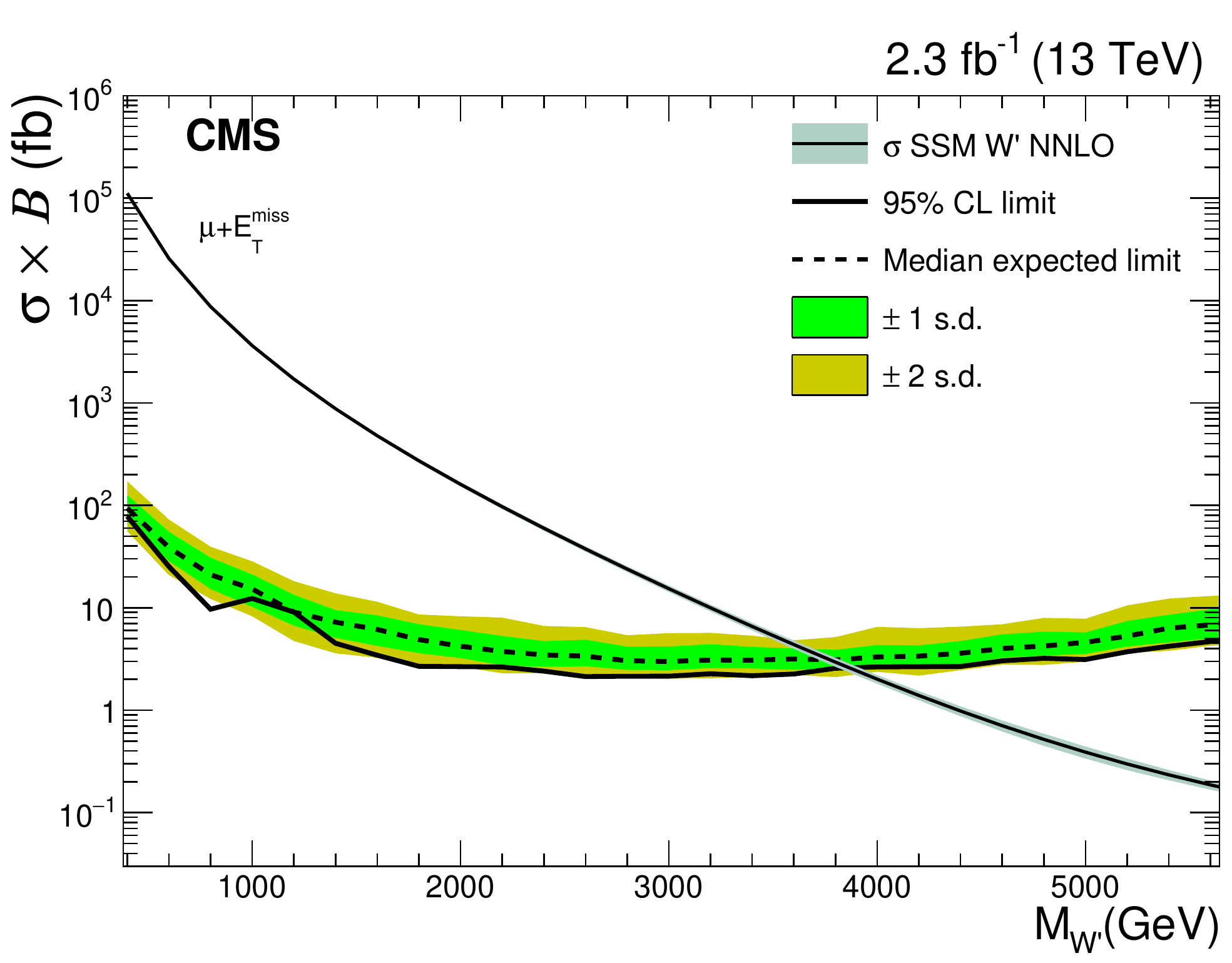}
 \caption{Expected and observed 95\% CL limits for the electron (\cmsLeft) and muon (\cmsRight)
decay channels. The expected (observed) limit is displayed as a dashed (solid) line and the associated inner (outer) bands represent the one (two) standard deviation (s.d.) uncertainties. The SSM \Wprime NNLO cross sections are displayed as a function of $M(\Wprime)$.
}
  \label{fig:limits-masswp}
\end{figure}

\begin{figure}[hbt]
\centering
 \includegraphics[width=\cmsFigWidth]{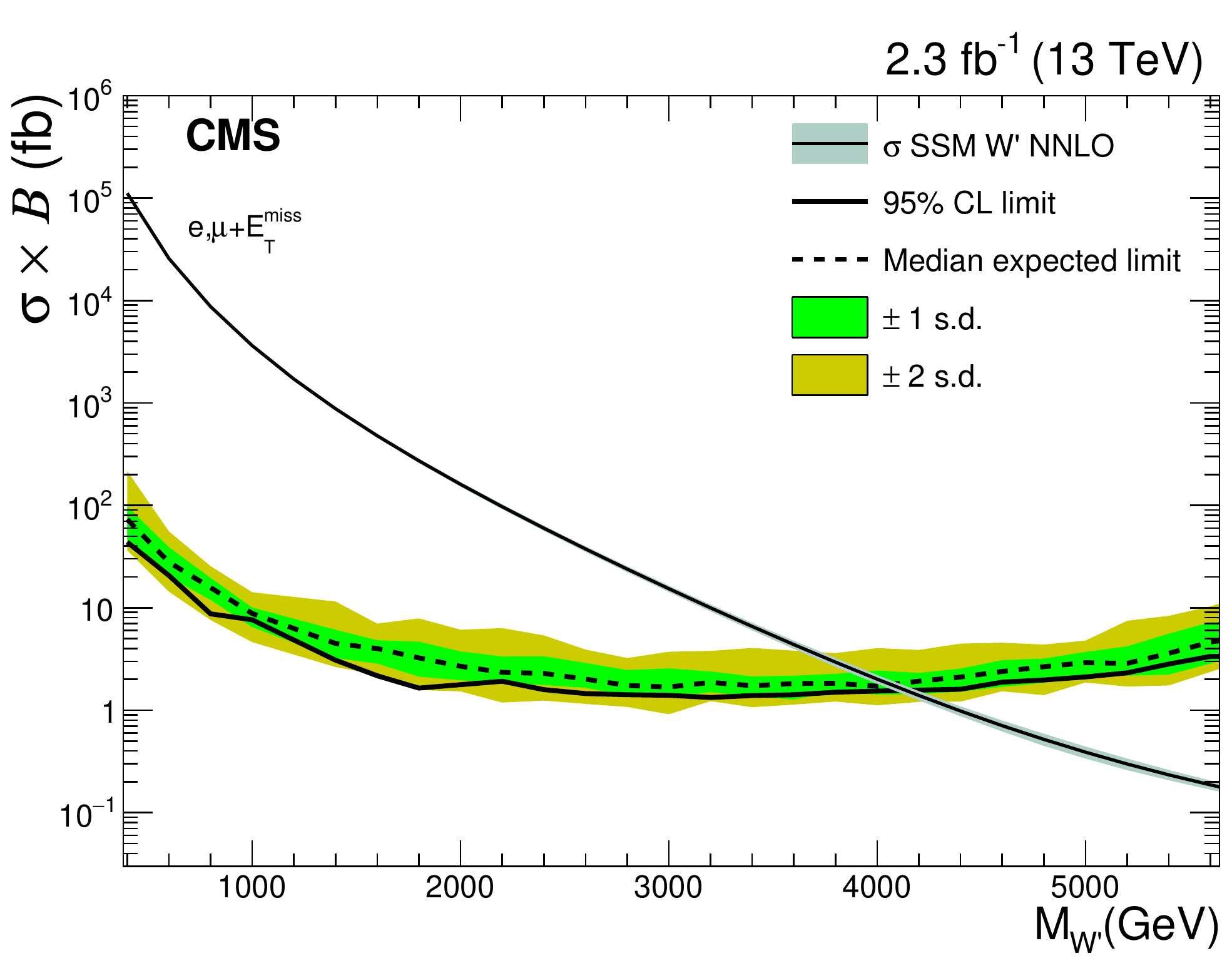}
 \caption{Expected and observed 95\% CL limits for the combination of the electron and muon decay channels. The expected (observed) limit is displayed as a dashed (solid) line and the associated inner (outer) bands represent the one (two) standard deviation (s.d.) uncertainties. The SSM \Wprime NNLO cross section is displayed as a function of $M(\Wprime)$.
}
  \label{fig:limits-combined}
\end{figure}

Expected and observed 95\% CL limits as a function of \Wprime mass are shown in Fig.~\ref{fig:limits-masswp} in the electron (\cmsLeft) and muon (\cmsRight) channels, for $M(\Wprime) > 400\GeV$.
The SSM \Wprime NNLO cross section as a function of the \Wprime mass
is also displayed, together with the uncertainty associated with the choice of PDF and $\alpha_S$, which is shown as a shaded band.
With the present data sample, SSM \Wprime resonances of masses less than 3.6\TeV (3.6\TeV expected) in the electron channel and less than 3.9\TeV (3.8\TeV expected) in the muon channel are excluded at 95\% CL.
These results provide tighter limits than those obtained from Run 1 data~\cite{EXO-12-060}.
The combination of the electron and muon channels,
 which have comparable sensitivity,
 improves the limit such that the production of SSM \Wprime bosons with masses below 4.1\TeV (4.0\TeV expected) are excluded at
95\% CL, as shown in Fig.~\ref{fig:limits-combined}. In making this combination, all systematic uncertainties that are common to both channels are
assumed to be fully correlated.

\subsection{Combination with Run 1 results}
\label{sec:results-combination}

\begin{figure}[hbtp]
\centering
 \includegraphics[width=0.49\textwidth]{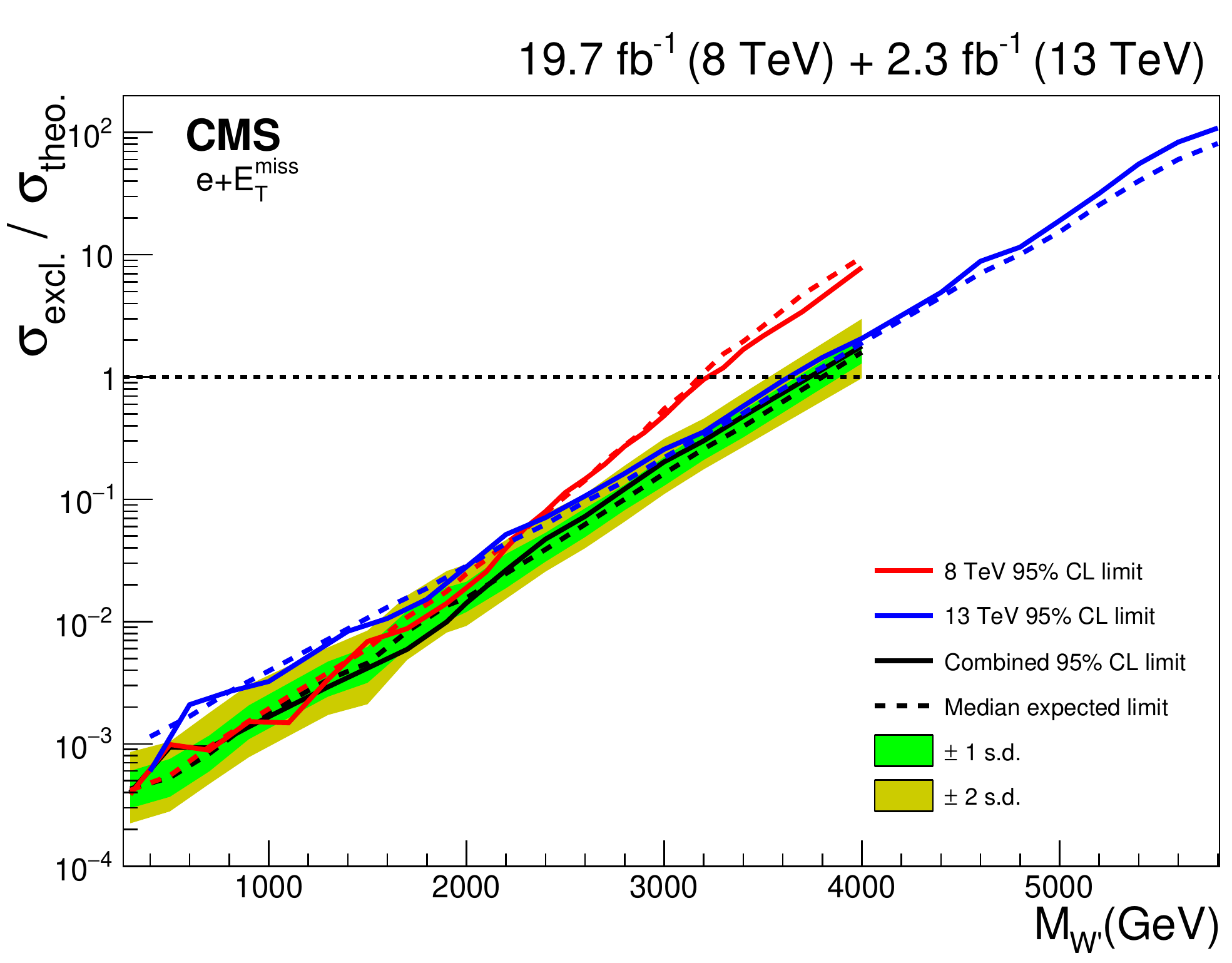}
 \includegraphics[width=0.49\textwidth]{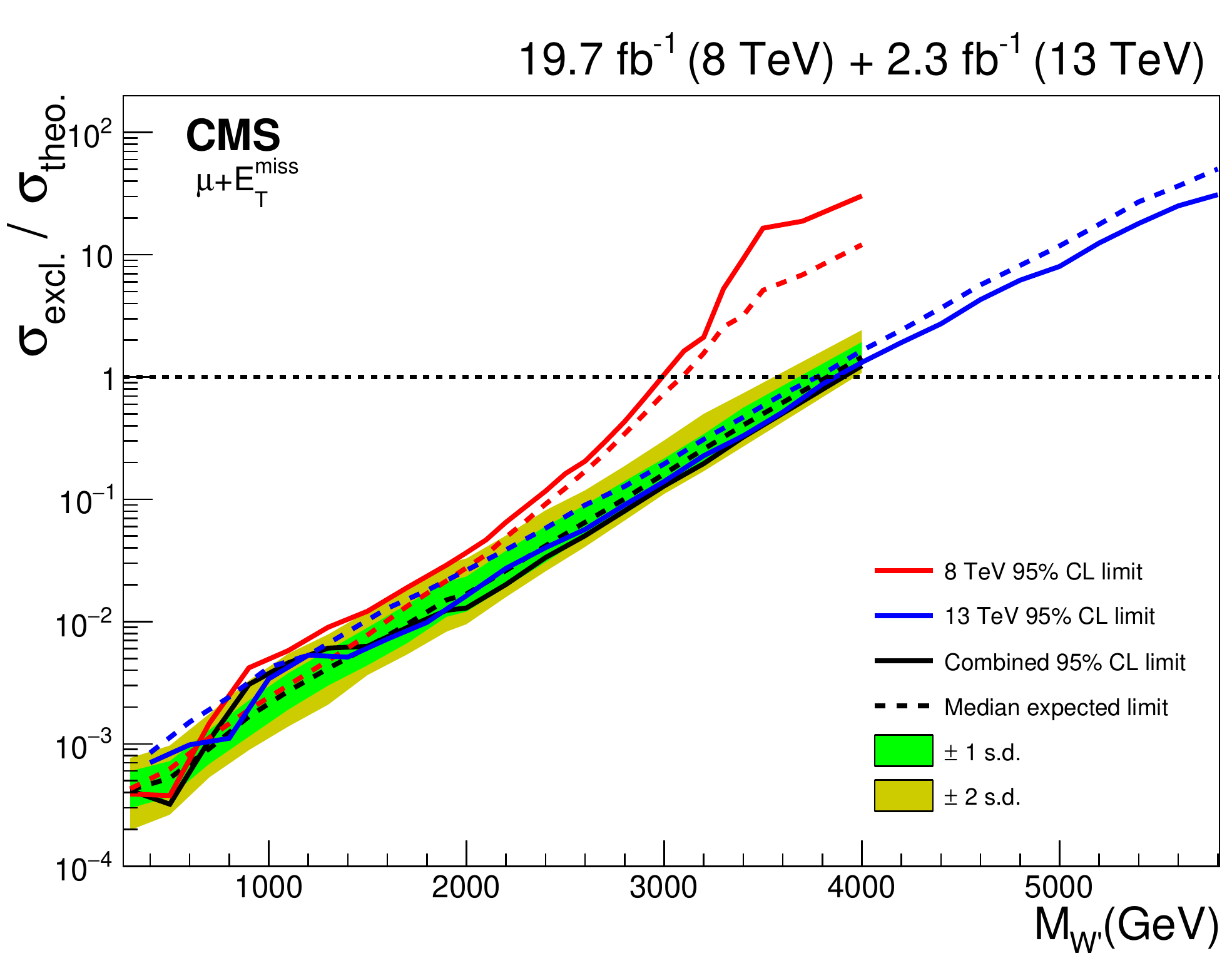}
 \caption{Ratio of the cross section lower limit to the theoretical cross section shown in red (lighter curves) for the 8\TeV data, blue (darker curves) for the 13\TeV data, and black for the combined data sets in the electron (\cmsLeft) and muon (\cmsRight) channels. The observed limits are shown with solid lines and expected limits with dashed lines. For the \Wprime mass range 400-4000\GeV, where the
combination is limited by the 8\TeV data set, the one- and two-standard deviation (s.d.) uncertainty bands for the combined limits are shown in green and yellow, respectively.
}
  \label{fig:energy-combined}
\end{figure}

A similar search for a \Wprime boson in the electron and muon channels was performed using Run 1 data at 8\TeV center-of-mass energy~\cite{EXO-12-060}.
These results can be combined with the present analysis using the prescription from
Ref.~\cite{Higgs_combi}.
The systematic uncertainties are assumed to be uncorrelated between Run 1 and Run 2.
The 95\% CL limits on the product $\sigma_{\Wprime} \mathcal{B}(\Wprime \to \ell \cPgn)$
derived from the combination of data at
$\sqrt{s} = 8$ and 13\TeV are presented in Fig.~\ref{fig:energy-combined} for
the electron (\cmsLeft) and the muon (\cmsRight) decay channels. In this case, the cross sections are presented relative to the
predicted NNLO cross section for the SSM \Wprime production at each center-of-mass energy.
The sensitivity to exclude high-mass \Wprime bosons is dominated by the data set at $\sqrt{s} = 13\TeV$,
and these data determine the limit exclusively for masses above 4 \TeV. For \Wprime masses below 2.2 \TeV, the higher integrated luminosity data set from the 8 \TeV Run still makes the biggest contribution to the sensitivity. Considering both data sets, SSM \Wprime bosons with masses less than 3.7 (3.9)\TeV are excluded in the electron (muon) channel. Combining both final state channels using the data at both center-of-mass energies the production of SSM \Wprime bosons with masses below 4.1\TeV is excluded at 95\%CL.

\subsection{Model-independent cross section limits}
\label{sec:results-xsec}

\begin{figure}[hbt]
\centering
  \includegraphics[width=0.45\textwidth]{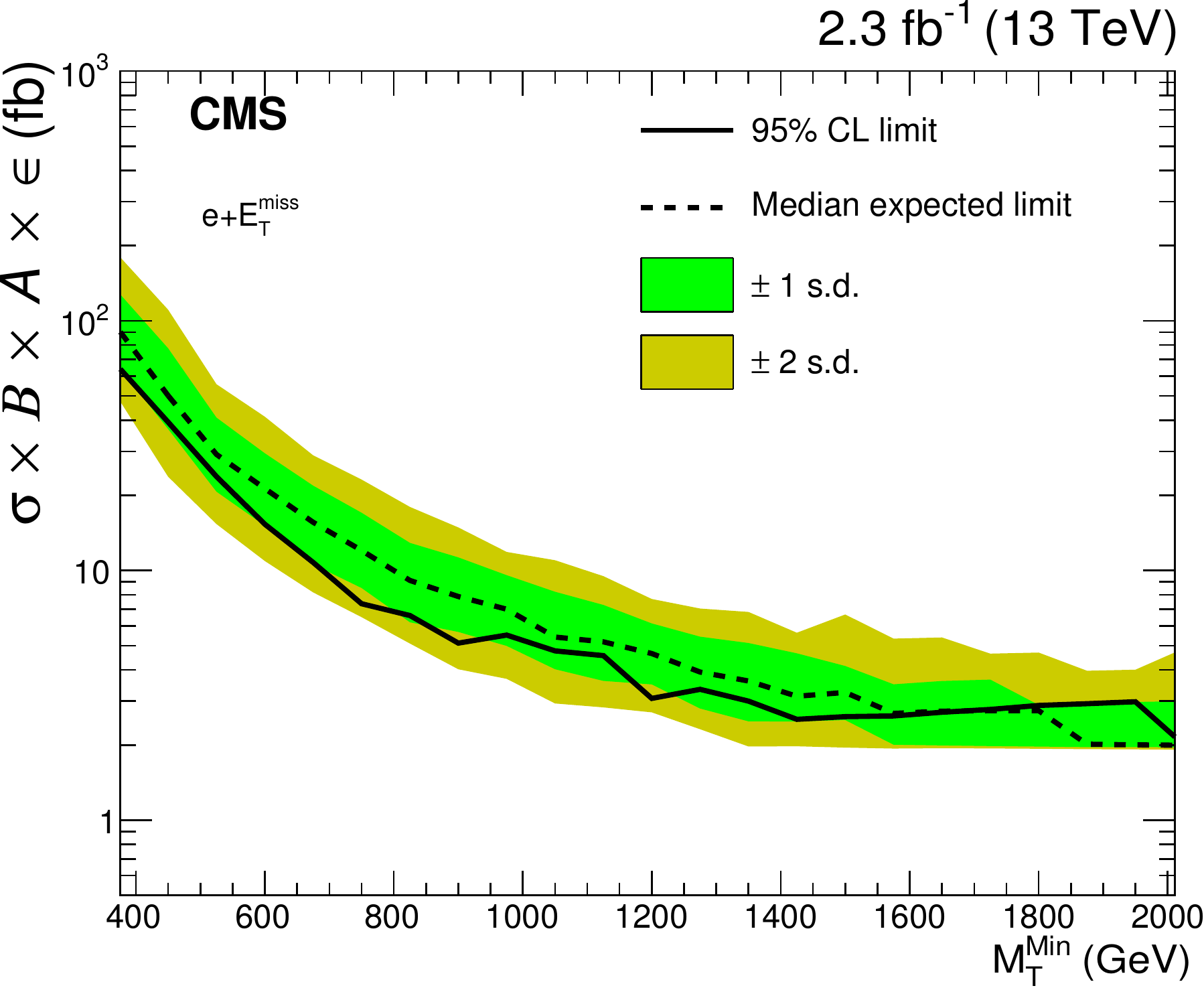}
  \includegraphics[width=0.45\textwidth]{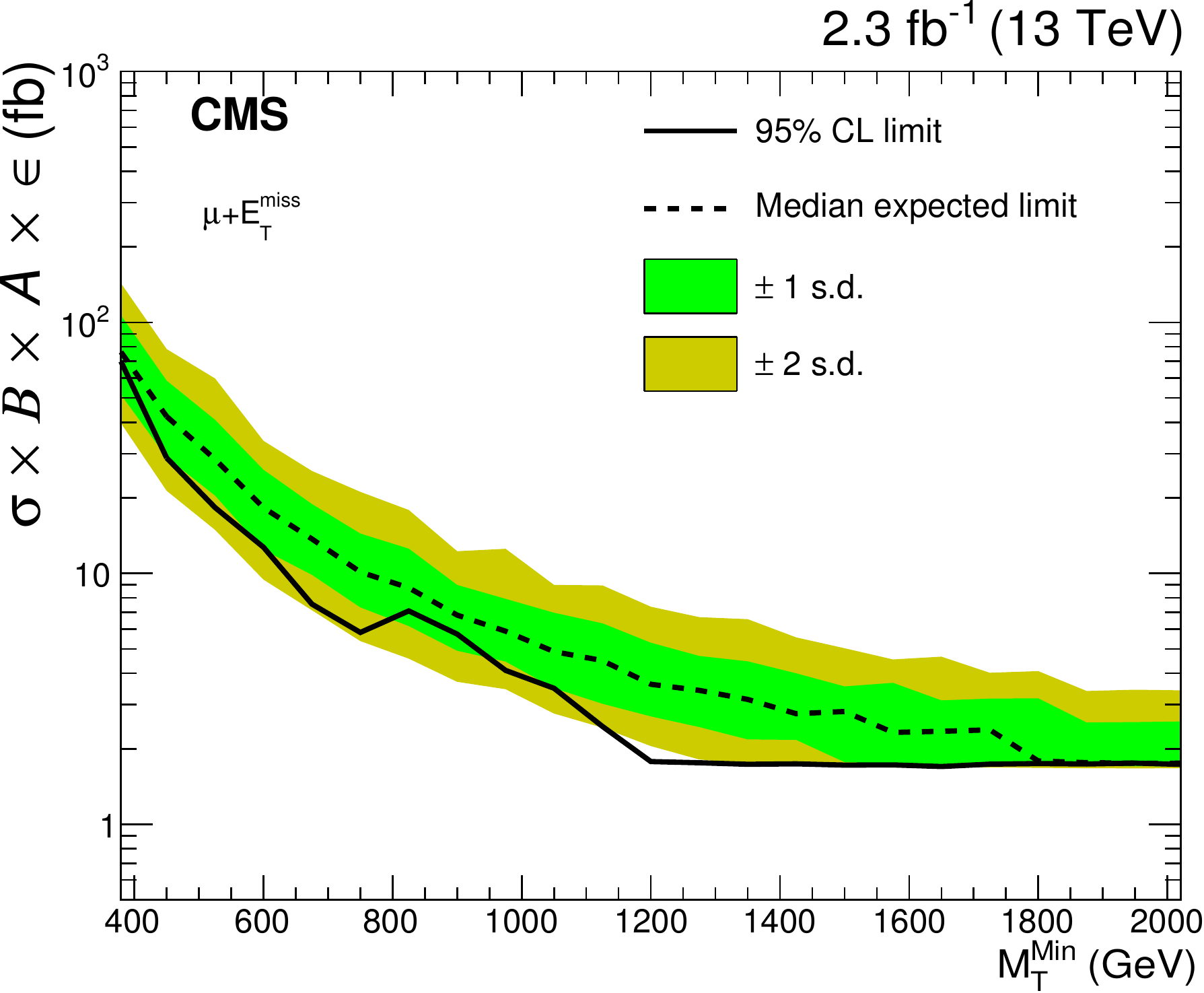}\\
  \includegraphics[width=0.45\textwidth]{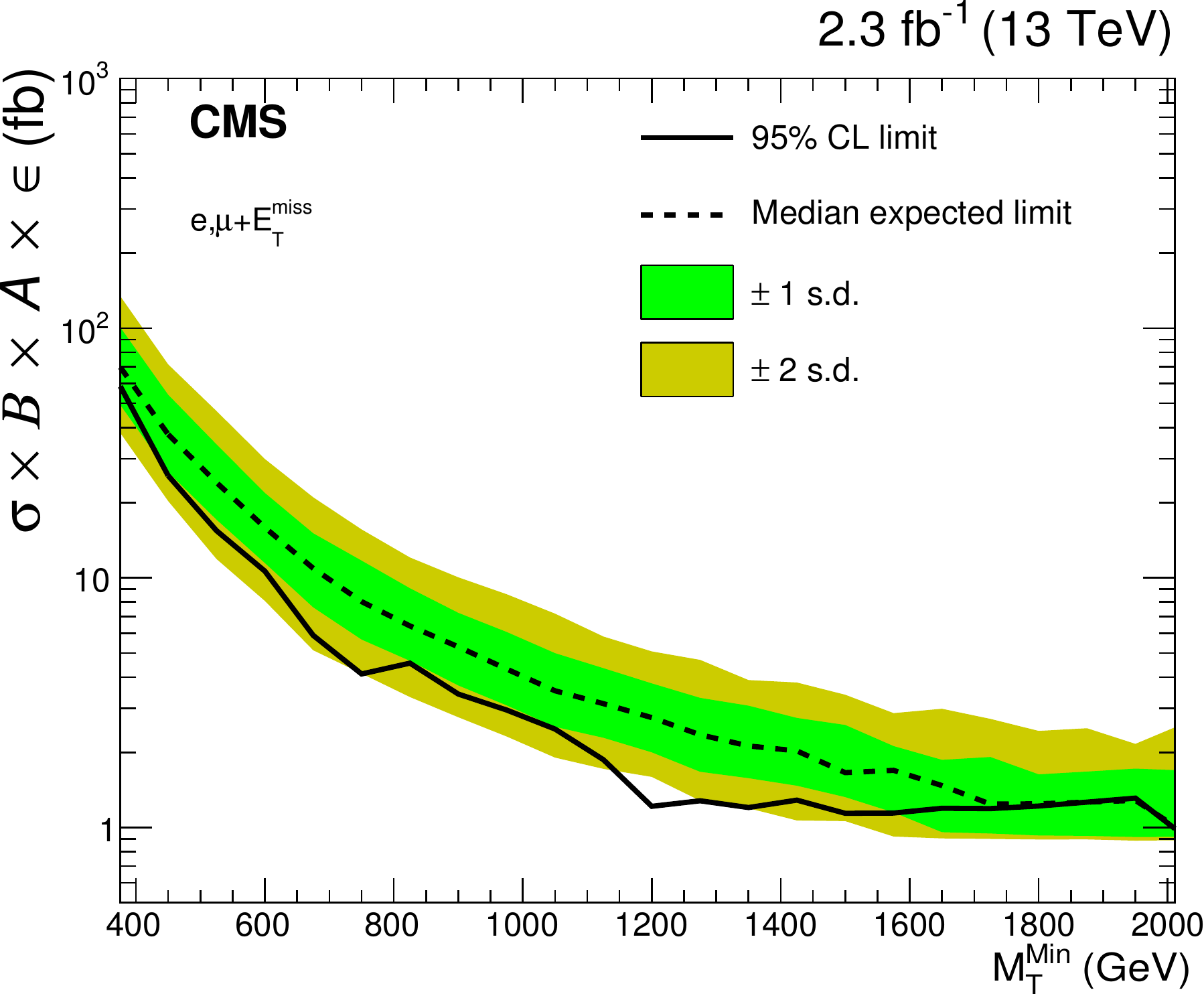}
 \caption{Expected (dashed line) and observed (solid line) 95\% CL limit for the model-independent cross section limit
    as a function of the lower \MT threshold, denoted \MTlower, for the electron channel (\cmsULeft), the muon channel (\cmsURight), and the combined channels (lower plot).
    The inner (outer) bands represent the one- (two-)standard deviation (s.d.) uncertainties.}
  \label{fig:MIlimits}
\end{figure}

A cross section limit that is independent of the \MT dependence expected in any given model is determined  by performing a single-bin counting experiment in a transverse mass range above a threshold, denoted \MTlower.
The results for the electron and muon channels are shown in Fig.~\ref{fig:MIlimits} along with the combination.
Values of the product of cross section and branching fractions above the solid curve are excluded.
The observed cross section limit includes the fiducial acceptance, $A$, defined by the lepton geometrical acceptance and the offline \pt thresholds (Section~\ref{sec:selection}), as well as detector effects and
kinematic selection (back-to-back topology), denoted as $\epsilon$. Both quantities are evaluated relative to events generated with a transverse mass above the \MTlower threshold.
The fiducial acceptance for very massive SSM W' bosons is of the order of 1, since the products of their decay are mainly emitted at very high angles relative to the beam direction.

In order to compare a specific new model to the given cross section limits, the effect of the threshold \MTlower on the signal acceptance has to be taken into account by
determining the ratio ($f_{\MT}$) of the number of events with $\MT > \MTlower$ to the number of events generated.
For the \MT range shown in Fig.~\ref{fig:MIlimits} the reconstruction efficiency is constant and the impact of the \MT resolution effect is negligible.
Therefore $f_{\MT}$ can be evaluated at generator level.
For lower \MT a very small ($<$1\%) difference is expected because of the single lepton trigger threshold (130\GeV for electrons, 50\GeV for muons).

A limit on the product of the cross section and branching fraction $(\sigma \, \mathcal{B} \, A \, \epsilon)_\text{excl}$ can be obtained by dividing the excluded cross section of the model-independent limit
$(\sigma \, \mathcal{B} \,  A \,  \epsilon)_\mathrm{MI}$ given in Fig.~\ref{fig:MIlimits}  by the calculated fraction $f_{\MT} (\MTlower)$:
\begin{equation*}
(\sigma \, \mathcal{B} \, A \,  \epsilon)_\text{excl} = \frac{(\sigma \, \mathcal{B} \, A \, \epsilon)_\mathrm{MI}(\MTlower)}{f_{\MT}(\MTlower)}.
\end{equation*}

Any deviation in the value of the product of the fiducial acceptance and signal efficiency of the new model from that applied to the \Wprime in Fig.~\ref{fig:MIlimits} would need to be taken into consideration. The latter has a value of 0.83$\pm$0.03, where the quoted uncertainty corresponds to the estimated variation as a function of \MTlower. For a predicted massive state decaying into two back-to-back leptons, thus having similar kinematic properties to those of a generic \Wprime boson, the deviation would be small and no additional correction would be required.

The validity of the model-independent limit procedure was checked by applying it
to an SSM \Wprime boson of 3.6\TeV mass and the results obtained are consistent with those presented in Section~\ref{sec:results-ssm} using the dedicated analysis.
It should be noted that this approach corresponds to a single-bin limit, which is expected to be slightly less sensitive than that obtained from a dedicated analysis
exploiting the full \MT shape.

\section{Summary}
\label{sec:summary}

A search has been performed for sequential standard model \Wprime bosons in final states containing a single energetic electron or muon and large missing transverse momentum,
using proton-proton collision data at $\sqrt{s}=13\TeV$, corresponding to an integrated luminosity of 2.3\fbinv.
No deviation from the background expectations has been observed and exclusion limits at 95\% confidence level have been extracted on the mass of a \Wprime boson. Masses below 3.6 (3.9)\TeV are excluded in the electron
(muon) decay channel analysis,
significantly improving upon the results obtained with the $\sqrt{s}=8\TeV$ data.
This search has been combined with the earlier one conducted at $8\TeV$, where the sensitivity of the search is dominated by the 13\TeV data, yielding a lower mass limit of 4.1\TeV for \Wprime bosons when
combining data from both decay channels and center-of-mass energies. Finally, generic limits on the production of  \Wprime resonances with the same leptonic final states have been obtained using a
model-independent approach.

\clearpage

\begin{acknowledgments}
We congratulate our colleagues in the CERN accelerator departments for the excellent performance of the LHC and thank the technical and administrative staffs at CERN and at other CMS institutes for their contributions to the success of the CMS effort. In addition, we gratefully acknowledge the computing centers and personnel of the Worldwide LHC Computing Grid for delivering so effectively the computing infrastructure essential to our analyses. Finally, we acknowledge the enduring support for the construction and operation of the LHC and the CMS detector provided by the following funding agencies: BMWFW and FWF (Austria); FNRS and FWO (Belgium); CNPq, CAPES, FAPERJ, and FAPESP (Brazil); MES (Bulgaria); CERN; CAS, MoST, and NSFC (China); COLCIENCIAS (Colombia); MSES and CSF (Croatia); RPF (Cyprus); SENESCYT (Ecuador); MoER, ERC IUT, and ERDF (Estonia); Academy of Finland, MEC, and HIP (Finland); CEA and CNRS/IN2P3 (France); BMBF, DFG, and HGF (Germany); GSRT (Greece); OTKA and NIH (Hungary); DAE and DST (India); IPM (Iran); SFI (Ireland); INFN (Italy); MSIP and NRF (Republic of Korea); LAS (Lithuania); MOE and UM (Malaysia); BUAP, CINVESTAV, CONACYT, LNS, SEP, and UASLP-FAI (Mexico); MBIE (New Zealand); PAEC (Pakistan); MSHE and NSC (Poland); FCT (Portugal); JINR (Dubna); MON, RosAtom, RAS, RFBR and RAEP (Russia); MESTD (Serbia); SEIDI and CPAN (Spain); Swiss Funding Agencies (Switzerland); MST (Taipei); ThEPCenter, IPST, STAR, and NSTDA (Thailand); TUBITAK and TAEK (Turkey); NASU and SFFR (Ukraine); STFC (United Kingdom); DOE and NSF (USA).

\hyphenation{Rachada-pisek} Individuals have received support from the Marie-Curie program and the European Research Council and EPLANET (European Union); the Leventis Foundation; the A. P. Sloan Foundation; the Alexander von Humboldt Foundation; the Belgian Federal Science Policy Office; the Fonds pour la Formation \`a la Recherche dans l'Industrie et dans l'Agriculture (FRIA-Belgium); the Agentschap voor Innovatie door Wetenschap en Technologie (IWT-Belgium); the Ministry of Education, Youth and Sports (MEYS) of the Czech Republic; the Council of Science and Industrial Research, India; the HOMING PLUS program of the Foundation for Polish Science, cofinanced from European Union, Regional Development Fund, the Mobility Plus program of the Ministry of Science and Higher Education, the National Science Center (Poland), contracts Harmonia 2014/14/M/ST2/00428, Opus 2014/13/B/ST2/02543, 2014/15/B/ST2/03998, and 2015/19/B/ST2/02861, Sonata-bis 2012/07/E/ST2/01406; the Thalis and Aristeia programs cofinanced by EU-ESF and the Greek NSRF; the National Priorities Research Program by Qatar National Research Fund; the Programa Clar\'in-COFUND del Principado de Asturias; the Rachadapisek Sompot Fund for Postdoctoral Fellowship, Chulalongkorn University and the Chulalongkorn Academic into Its 2nd Century Project Advancement Project (Thailand); and the Welch Foundation, contract C-1845.

\end{acknowledgments}

\clearpage
\bibliography{auto_generated}

\cleardoublepage \appendix\section{The CMS Collaboration \label{app:collab}}\begin{sloppypar}\hyphenpenalty=5000\widowpenalty=500\clubpenalty=5000\textbf{Yerevan Physics Institute,  Yerevan,  Armenia}\\*[0pt]
V.~Khachatryan, A.M.~Sirunyan, A.~Tumasyan
\vskip\cmsinstskip
\textbf{Institut f\"{u}r Hochenergiephysik,  Wien,  Austria}\\*[0pt]
W.~Adam, E.~Asilar, T.~Bergauer, J.~Brandstetter, E.~Brondolin, M.~Dragicevic, J.~Er\"{o}, M.~Flechl, M.~Friedl, R.~Fr\"{u}hwirth\cmsAuthorMark{1}, V.M.~Ghete, C.~Hartl, N.~H\"{o}rmann, J.~Hrubec, M.~Jeitler\cmsAuthorMark{1}, A.~K\"{o}nig, I.~Kr\"{a}tschmer, D.~Liko, T.~Matsushita, I.~Mikulec, D.~Rabady, N.~Rad, B.~Rahbaran, H.~Rohringer, J.~Schieck\cmsAuthorMark{1}, J.~Strauss, W.~Waltenberger, C.-E.~Wulz\cmsAuthorMark{1}
% ...
\vskip\cmsinstskip
% ...
1:~~Also at Vienna University of Technology, Vienna, Austria\\
% ...

\end{sloppypar}
\end{document}